\documentclass[12pt]{article}

\usepackage{graphicx}
\usepackage{amsmath}
\usepackage{amsfonts}
\usepackage{amssymb}
\usepackage{verbatim}
\usepackage{epsfig}

\renewcommand{\arraystretch}{1.3}

\makeatletter
\newdimen\normalarrayskip              
\newdimen\minarrayskip                 
\normalarrayskip\baselineskip
\minarrayskip\jot
\newif\ifold             \oldtrue            \def\new{\oldfalse}
\def\arraymode{\ifold\relax\else\displaystyle\fi} 
\def\eqnumphantom{\phantom{(\theequation)}}     
\def\@arrayskip{\ifold\baselineskip\z@\lineskip\z@
     \else
     \baselineskip\minarrayskip\lineskip2\minarrayskip\fi}
\def\@arrayclassz{\ifcase \@lastchclass \@acolampacol \or
\@ampacol \or \or \or \@addamp \or
   \@acolampacol \or \@firstampfalse \@acol \fi
\edef\@preamble{\@preamble
  \ifcase \@chnum
     \hfil$\relax\arraymode\@sharp$\hfil
     \or $\relax\arraymode\@sharp$\hfil
     \or \hfil$\relax\arraymode\@sharp$\fi}}
\def\@array[#1]#2{\setbox\@arstrutbox=\hbox{\vrule
     height\arraystretch \ht\strutbox
     depth\arraystretch \dp\strutbox
     width\z@}\@mkpream{#2}\edef\@preamble{\halign
\noexpand\@halignto
\bgroup \tabskip\z@ \@arstrut \@preamble \tabskip\z@ \cr}%
\let\@startpbox\@@startpbox \let\@endpbox\@@endpbox
  \if #1t\vtop \else \if#1b\vbox \else \vcenter \fi\fi
  \bgroup \let\par\relax
  \let\@sharp##\let\protect\relax
  \@arrayskip\@preamble}
\def\eqnarray{\stepcounter{equation}%
              \let\@currentlabel=\theequation
              \global\@eqnswtrue
              \global\@eqcnt\z@
              \tabskip\@centering
              \let\\=\@eqncr

 \halign to \displaywidth\bgroup
    \eqnumphantom\@eqnsel\hskip\@centering
    $\displaystyle \tabskip\z@ {##}$%
    \global\@eqcnt\@ne \hskip 2\arraycolsep
         $\displaystyle\arraymode{##}$\hfil
    \global\@eqcnt\tw@ \hskip 2\arraycolsep
         $\displaystyle\tabskip\z@{##}$\hfil
         \tabskip\@centering
    &{##}\tabskip\z@\cr}
\newfont{\hr}{msbm10}
\newfont{\ams}{msam10}

\voffset= - 1.1in
\hoffset= - 0.9in         

\def\beq{\begin{equation}}
\def\eeq{\end{equation}}
\def\ba{\beq\new\begin{array}{c}}
\def\ea{\end{array}\eeq}
\def\be{\ba}
\def\ee{\ea}

\def\N2{${\cal N}=2$}

\def\1N{${\cal N}=1$}
\def\4N{${\cal N}=4$}

\def\p{\partial}

\newdimen\linethick  \linethick=0.4pt
\newdimen\hboxitspace    \hboxitspace=5pt
\newdimen\vboxitspace    \vboxitspace=5pt

\def\fr#1{%
\beq\new
\vcenter{
\hrule height\linethick
          \hbox{\vrule width\linethick
                \kern\hboxitspace
                \vbox{\kern\vboxitspace
                      \hbox{$\begin{array}{c}\displaystyle#1
         \end{array}$}%
                      \kern\vboxitspace}%
                \kern\hboxitspace
                \vrule width\linethick}%
          \hrule height\linethick}%
\eeq}

\def\p{\partial}

\def\p{\partial}

\def\langle{\left<}
\def\rangle{\right>}

\begin{document}
\begin{flushright}
INR-TH-2012-72
\end{flushright}
\vspace{10pt}
\begin{center}
  {\LARGE \bf Pseudo-conformal Universe: late-time  \\[0.3cm]
contraction and generation of tensor modes} \\
\vspace{20pt}
S.~Mironov\\
\vspace{20pt}
\textit{
Institute for Nuclear Research of
         the Russian Academy of Sciences,\\  60th October Anniversary
  Prospect, 7a, 117312 Moscow, Russia;}\\
\vspace{10pt}
\textit{
Institute of Theoretical and Experimental Physics\\ Bolshaya Cheremushkinskaya 25,
117218, Moscow, Russia
}

\end{center}

    \vspace{5pt}

\begin{abstract}

We consider a bouncing Universe model which explains the flatness of the primordial scalar spectrum via complex scalar field that rolls down its negative quartic potential and dominates in the Universe. We show that in this model, there exists a rapid contraction regime of classical evolution. We calculate the power spectrum of tensor modes in this scenario. We find that it is blue and its amplitude is typically small, leading to mild constraints on the parameters of the model.

\end{abstract}
\section{Introduction}

Inflation is the best studied and very plausible scenario for the cosmological epoch preceding the hot Big Bang era. Nevertheless, it is of interest to consider alternative scenarios. One of them is Ekpyrosis \cite{ekpyrosis,ekpnew}, understood in a broad sense as a period of slow contraction with equation of state $p>\rho$. Its particular version, pseudo-conformal Universe \cite{hinterbicherkhouri,hinterbicherkhourilr} makes use of a nearly conformal evolution to generate scalar perturbations with nearly flat power spectrum, thus building upon the idea that the flat spectrum may be due to conformal invariance \cite{confinfl1,confinfl2,confinfl3}. The adiabatic perturbations generated by conformal mechanisms have interesting non-linear properties, such as non-Gaussianity and statistical anisotropy \cite{confnew}.

The concrete example of the pseudo-conformal Universe \cite{hinterbicherkhouri} is based on the Einstein gravity interacting minimally with a complex scalar field with negative quartic potential. The action is (signature $(-+++)$)
 \be
S=-\frac{M_{pl}^2}{16\pi}\int d^4x\sqrt{-g}R+\int d^4x\sqrt{-g}\left[\frac{1}{2}g_{\mu\nu}\p_{\mu}\phi\p_{\nu}\phi^*+\frac{\lambda}{4}|\phi|^4 \right] \; .
\nonumber
\ee
The contracting epoch is assumed to begin with the asymptotically Minkowski space-time and $\phi=0$. Since the potential is negative, this state is unstable, and under the assumption of the homogeneous background evolution, the field rolls down as follows, 
\be
\phi=\frac{\sqrt{2}}{\sqrt{\lambda}(-t)}~,~~~~~~~t<0~,~~~~~~~|t|\rightarrow\infty \; .
\label{phi1}
\ee
At this stage perturbations of the phase $\mbox{Arg}~\phi$ are generated, which, to the linear order, automatically have flat power spectrum \cite{confinfl2}. These entropy perturbations are assumed to get reprocessed into adiabatic ones by one or another mechanism, see, e.g., Refs. \cite{curv,moddec,wandbrand}. It has been pointed out \cite{hinterbicherkhouri} that tensor perturbations are generated as well, and that they have blue power spectrum. The latter is similar to the tensor spectra generated in other ekpyrotic models \cite{gravwave}.

In this paper our focus is on the late-time contraction in this model. The classical behavior turns out to be rather interesting: the initial slow contraction with the effective equation of state $p\gg\rho$ is followed by the rapid contraction regime in which the equation of state asymptotes to $p=\rho$. As in all ekpyrosis-like models, one assumes that the latter regime ends up in a bounce and defrosting \cite{def} whose mechanism we do not specify.

We also consider the generation of tensor modes, both at slow and rapid contraction stages. We find that their power spectrum is always blue, and the amplitude is small enough, so that no strong constraints on the parameters of the model can be placed. Thus, the model is safe from this viewpoint.

This paper is organized as follows. In Section 2 we consider the classical contraction. In Section 3 we study the tensor perturbations. We conclude in Section 4.

\section{Classical solution}

The homogeneous contracting Universe is governed by the equation of motion for the field $\phi$,
\be
\ddot{\phi}+3H\dot{\phi}=\lambda\phi^3 \; ,
\label{urdv}
\ee
and the Friedmann equation
\be
H^2=\frac{8\pi}{3M_{pl}^2}\left(\frac{1}{2}\dot{|\phi|}^2-\frac{\lambda}{4}|\phi|^4\right) \; .
\label{urfr}
\ee
At early times, $t\rightarrow -\infty$, one neglects the Hubble friction term in the first equation and obtains the solution (\ref{phi1}), where we take $\phi$ real. To this order, the energy density of the field $\phi$ vanishes. The Hubble parameter and then the scale factor can be obtained from the Raychaudhuri equation
\be
\frac{2\ddot{a}}{a}+\left(\frac{\dot{a}}{a}\right)^2=-\frac{8\pi}{M_{pl}^2}\left(\frac{\dot{\phi}^2}{2}+\frac{\lambda\phi^4}{4}\right) \; .
\nonumber
\ee
One finds
\be
H=-\frac{8\pi}{3\lambda M_{pl}^2}\frac{1}{(-t)^3} \; .
\label{hfirst}
\ee
This corresponds to the equation of state
\be
\frac{p}{\rho}\rightarrow\infty \; .
\nonumber
\ee
From eq.(\ref{hfirst}) we get the scale factor
\be
a=a_{**}\left(1-\frac{4\pi}{3\lambda M_{pl}^2}\frac{1}{t^2}\right) \; ,
\label{afirst}
\ee
where we keep the overall normalization factor $a_{**}$ as a free parameter.
So, the Universe contracts slowly at early times.

We are mainly interested in the solution at late times. For large $H$ we neglect the scalar potential and solve the simplified equation of motion and Friedmann equation:
\be
\ddot{\phi}+3H\dot{\phi}=0 \; ,
\nonumber
\ee
\be
H=-\frac{2\sqrt{\pi}}{\sqrt{3}M_{pl}}\dot{\phi} \; .
\label{H2}
\ee
Upon substituting $H$ into the first equation we get
\be
\ddot{\phi}=\frac{\sqrt{12\pi}}{M_{pl}}\dot{\phi}^2 \; .
\nonumber
\ee
The solution is
\be
\phi=-\frac{M_{pl}}{\sqrt{12\pi}}~\ln(\mu(t_*-t)) \; ,
\label{phi2}
\ee
where $t_*$ and $\mu$ are to be determined by matching this solution to eq. (\ref{phi1}).
We can indeed neglect the scalar potential in eqs. (\ref{urdv}), (\ref{urfr}), provided that $\lambda\phi^4\ll \dot{\phi}^2$ and $\lambda\phi^3\ll\ddot{\phi}$. Using eq. (\ref{phi2}) we see that these conditions are satisfied at late times, such that
\be
t_*-t \ll \sqrt{\frac{12\pi}{\lambda}}\frac{1}{M_{pl}|\ln(\mu(t_*-t))|^{3/2}}
\label{ineq}
\ee
Note that for small $\lambda$, there exists a time interval where the evolution is still classical, $(t_*-t)\gg M_{pl}^{-1}$, while the inequality (\ref{ineq}) is satisfied.

It follows from eqs. (\ref{H2}) and (\ref{phi2}) that at late times, the Hubble parameter and scale factor are
\be
H=-\frac{1}{3(t_*-t)} \; ,
\label{Hsecond}
\ee
\be
a=a_*' (t_*-t)^{1/3} \; ,
\label{asecond}
\ee
where $a_*'$ is yet undetermined parameter.
They correspond to the equation of state
\be
p=\rho \; .
\nonumber
\ee

Let us match the two regimes and express the coefficients $t_*$,$a_*'$ and $\mu$ in terms of $a_{**}$.
For estimates, we require that $\phi$, $a$ and their time derivatives are equal for the solutions (\ref{phi1}), (\ref{afirst}) and (\ref{phi2}), (\ref{asecond}) at some time $t_{M}$. We write

\begin{align}
\phi:~~~~& \frac{\sqrt{2}}{\sqrt{\lambda}(-t_M)}=-\frac{M_{pl}}{\sqrt{12\pi}}\ln\left(\mu(t_*-t_M)\right) \; ,
\label{shiv1}\\
\dot{\phi}:~~~~& \frac{\sqrt{2}}{\sqrt{\lambda}(-t_M)^2}=\frac{1}{\sqrt{12\pi}}\frac{M_{pl}}{t_*-t_M} \; ,\\
a:~~~~& a_{**}=a_*'(t_*-t_M)^{1/3} \; ,\\
H:~~~~& \frac{8\pi}{3\lambda M_{pl}^2(-t_M)^3}=\frac{1}{3(t_*-t_M)} \; .
\label{shiv4}
\end{align}
As a cross check, the potential and Hubble friction terms in eq. (\ref{urdv}) are of the same order at time $t_{M}$:~$\lambda\phi^3 \sim 3H\dot{\phi}$. Likewise, the kinetic and potential terms in the Friedmann equation (\ref{urfr}) are of the same order at that time, $\dot{\phi}^2\sim \lambda\phi^4$.

\begin{figure}
\raisebox{-1mm}{ \includegraphics[height=100mm]{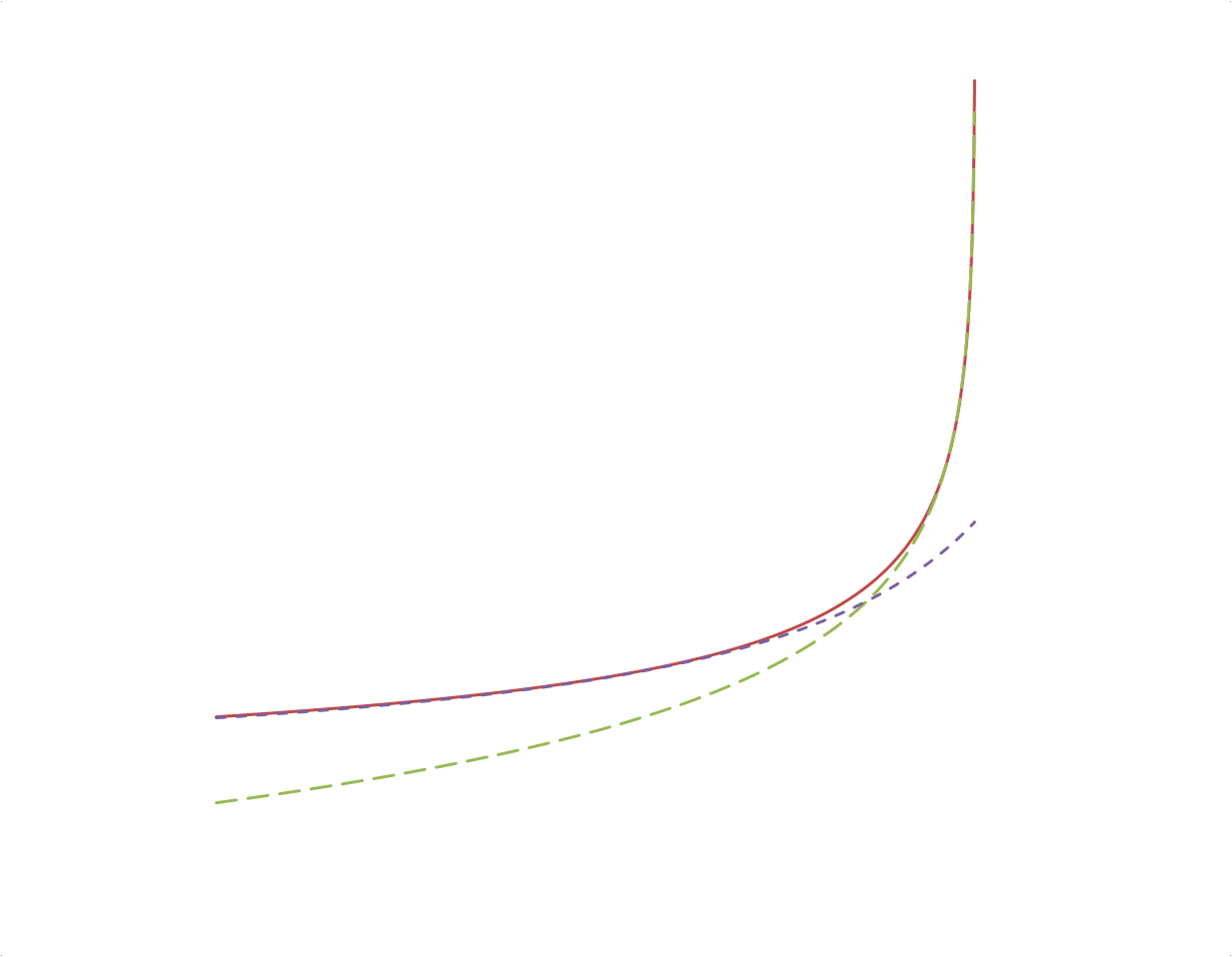}
\put(0,40){\line(-1,0){350}}
\put(0,40){\line(0,1){230}}
\put(10,245){$\phi$}
\put(-330,20){$t$}}
\caption{An example of numerical solution to eqs. (\ref{urdv}), (\ref{urfr}) (solid line). Short dashed line is $\sqrt{2}/(-\sqrt{\lambda}t)$. Long dashed line is  $-\frac{M_{pl}}{\sqrt{12\pi}}\ln(\mu(t_*-t))$.}
\label{f1}
\end{figure}
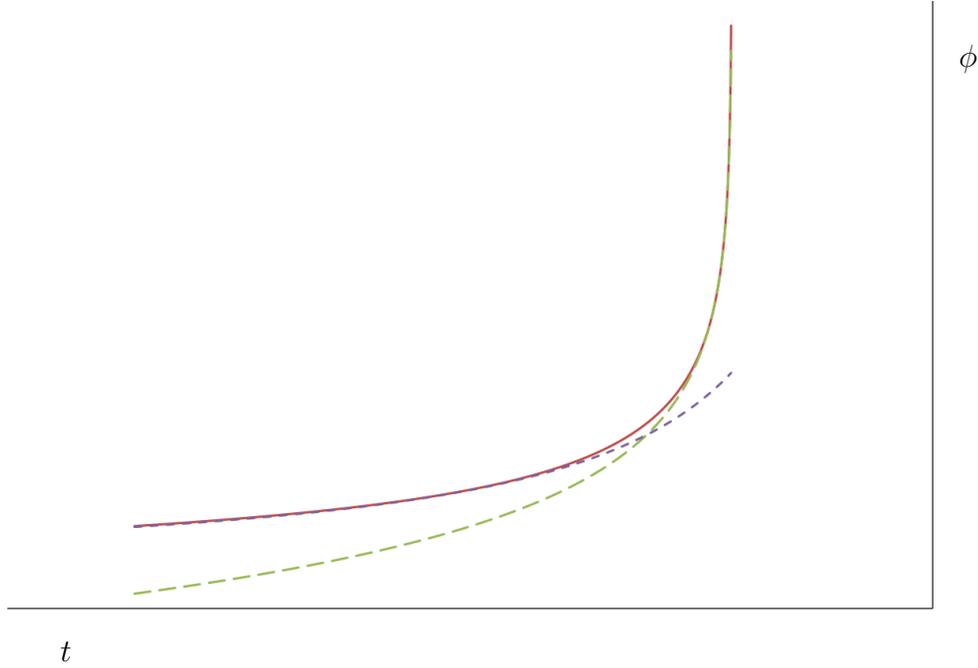

By solving eqs. (\ref{shiv1})~--~(\ref{shiv4}), we obtain the time of the transition from slow to fast contraction,
\be
t_M=-\sqrt{\frac{8\pi}{3}}\frac{1}{\sqrt{\lambda}M_{pl}} \; .
\label{sh1}
\ee
Of course, if the bounce happens at $t<t_M$, the fast contraction regime does not occur at all. Assuming that this is not the case, we find the parameters of the fast contraction stage
\be
\mu\sim
\sqrt{\lambda}M_{pl}
\label{sh2}
\ee
\be
t_{*}=-\sqrt{\frac{32\pi}{27}}\frac{1}{\sqrt{\lambda}M_{pl}} \; ,
\label{sh3}
\ee
\be
a_{*}'\simeq a_{**}\lambda^{1/6}M_{pl}^{1/3} \; .
\label{sh4}
\ee
An example of numerical solution is shown in Fig. \ref{f1}. One observes that the two regimes (\ref{phi1}) and (\ref{phi2}) indeed nicely match at intermediate time.

In what follows we use the solution in the second regime, written in terms of conformal time (in the first regime conformal time equals $t/a_{**}$):
\be
a=a_* (\eta_*-\eta)^{1/2} \; .
\label{acsecond}
\ee
Making use of eqs. (\ref{sh3}), (\ref{sh4}) and requiring that $\eta(t_{M})=t_{M}/a_{**}$, we get

\be
a_*^2\simeq a_{**}^3M_{pl}\sqrt{\lambda} \; ,
\label{sh5}
\ee

\be
\eta_* \sim \left(\frac{a_{**}}{a_{*}}\right)^2 \; .
\label{sh6}
\ee

To end up this Section, we express the parameters of our solutions in terms of physical quantities.
We consider the simpliest possibility that the contracting stage terminates abruptly by the bounce and instantaneous defrosting. We therefore assume that both the scale factor and energy density (and hence $|H|$) coincide at the end of the contracting stage and at the beginning of the hot epoch. At the beginning of the hot epoch just after the bounce we have
\be
H_b=\frac{T_{d}^{2}}{M_{pl}^{*}}
\label{Hubblerad}
\ee
\be
\frac{a_{b}}{a_{T}}=\frac{T}{T_{d}}\left(\frac{g_{*T}}{g_{*d}}\right)^{1/3}
\label{ecl}
\ee
where $T_{d}$ is the defrosting temperature, $M_{pl}^*=M_{pl}^*(T_{d})=\frac{M_{pl}}{1.66\sqrt{g_{*d}}}$ and $T$ refers to any moment of time at the radiation dominated epoch.
On the other hand, just before the bounce the Hubble parameter and scale factor are given by
\be
H_b=-\frac{1}{2a_*(\eta_*-\eta_b)^{3/2}} \; ,
\label{Hbreh}
\ee
\be
a_b=a_* (\eta_*-\eta_b)^{1/2} \; .
\nonumber
\ee
We thus obtain
\be
\eta_*-\eta_b=\frac{1}{2a_b}\frac{M_{pl}^*}{T_{d}^2}=\frac{M_{pl}^*g_{*d}^{1/3}}{2a_TTT_{d}g_{*T}^{1/3}} \; ,
\label{razet}
\ee
\be
a_*=\left(\frac{2a_T^3T^3}{T_{d}M_{pl}}\frac{1.66 g_{*T}}{\sqrt{g_{*d}}}\right)^{1/2} \; .
\label{a*phys}
\ee
Finally, the scale factor at the beginning of contraction is
\be
a_{**}=\left(\frac{2a_T^3 T^3}{T_{d}M_{pl}^2\sqrt{\lambda}}\frac{1.66 g_{*T}}{\sqrt{g_{*d}}}\right)^{1/3} \; .
\label{a**phys}
\ee
Let us now turn to the calculation of the power spectrum of tensor perturbations.

\section{Tensor perturbations}

In this Section we work in terms of the canonically normalized field 
\be
\varphi=\sqrt{\frac{M_{pl}^2}{32\pi}}h \; ,
\label{svuaz}
\ee
where $h$ is the shorthand for the transverse traceless component of the metric perturbation. The field equation has the following form in conformal time:
\be
\varphi''+2\frac{a'}{a}\varphi'+k^2\varphi=0 \; .
\label{oburvo}
\ee
In terms of the variable $\chi=a\varphi$ this equation reads:
\be
\chi''-\frac{a''}{a}\chi+\frac{k^2}{a^2}\chi=0 \; .
\label{urvoz}
\ee
In what follows we consider two regimes separately: early times, when the scale factor $a$ is varying slowly; and late times, when it evolves as $a\propto(t_*-t)^{1/3}$. The former regime is relevant for low momentum modes which exit the horizon early, while the latter is important for high momentum modes.

We begin with the modes of low momenta. In the slow contraction regime, the scale factor behaves as follows:
\be
a=a_{**}\left(1-\frac{c}{\eta^2}\right) ~~~~~ c=\frac{4\pi}{3\lambda M_{pl}^2a_{**}^2} \; .
\nonumber
\ee
The low momentum modes are those which exit the horizon, $k/a\sim H$, well before time $t_M$. Making use of eqs. (\ref{hfirst}) and (\ref{sh1}), we obtain for these modes
\be
k\ll k_c=\sqrt{\lambda}M_{pl}a_{**}=M_{pl}^{2/3}\lambda^{1/3}a_*^{2/3} \; .
\label{midm}
\ee
Note that for these modes one has $k\eta_{\times}\ll 1$, where $\eta_{\times}$ is the conformal time of horizon exit.

In the slow contraction regime, eq. (\ref{urvoz}) has the following explicit form:
\be
\chi''+\frac{6c}{\eta^4}\chi+k^2\chi=0 \; .
\label{eqlt}
\ee
We assume that the field is in its vacuum state at past infinity and hence has the standard asymptotics $\chi^{(\pm)}=e^{\pm ik\eta}$ at large $(-\eta)$. After crossing out the horizon, but before the onset of rapid contraction, the solution is
\be
\chi=C_1a+C_2a\int\frac{dt}{a^3}=C_1a_{**}\left(1-\frac{c}{\eta^2}\right)+\frac{C_2}{a_{**}}\left(\eta-\frac{3c}{\eta}\right) \; .
\label{zagor}
\ee
Now we match the two asymptotics and find $\chi^{(+)}$ and $\chi^{(-)}$ in the superhorizon regime.
To this end we solve eq. (\ref{eqlt}) perturbatively near horizon crossing. In the slow contraction regime we have $\frac{c}{\eta^2}\ll 1$, and the horizon crossing occurs when $k^2\sim\frac{c}{\eta^4}$. Hence, there are two small parameters around the horizon crossing: $k^2\eta^2,~\frac{c}{\eta^2}\ll 1$, and the two linear independent solutions to eq. (\ref{eqlt}) can be found as series in these parameters:
\be
\chi_{(1)}=1-\frac{c}{\eta^2}-\frac{1}{2}k^2\eta^2+...~~ \; ,
\\
\chi_{(2)}=\eta-\frac{3c}{\eta}-\frac{1}{6}k^2\eta^3+...~~ \; .
\label{asr}
\ee
Before the horizon  crossing, when $\frac{c}{\eta^2}\ll k^2\eta^2\ll 1$, these solutions match to $\chi_{(1)}=\cos(k\eta)$ and $\chi_{(2)}=\sin(k\eta)/k$. Since $\chi^{(+)}\sim e^{ik\eta}$, we get
\be
\chi^{(+)}=\chi_{(1)}+ik\chi_{(2)} \; .
\nonumber
\ee
After horizon crossing, when $k^2\eta^2\ll \frac{c}{\eta^2}\ll 1$, this solution reads
\be
\chi^{(+)}=\left(1-\frac{c}{\eta^2}\right)+ik\left(\eta-\frac{3c}{\eta}\right) \; ,
\label{finsol0}
\ee
which coincides with eq. (\ref{zagor}) with $C_1=a_{**}^{-1}$, $C_2=ika_{**}$. So, the low momentum mode after horizon exit is
\be
\varphi^{(+)}=\frac{1}{(2\pi)^{3/2}\sqrt{2k}}\left(\frac{1}{a_{**}}+ika_{**}\int\limits_{\eta_{\times}}^{\eta}\frac{d\eta}{a^2(\eta)}\right) \; ,
\label{finsol}
\ee
where we inserted the standard normalization factor and made use of the fact that $k\eta_{\times}\ll 1$. We cross check this calculation in Appendix.

The result (\ref{finsol}) is valid also at late times, when the Universe contracts rapidly.
Making use of eqs. (\ref{afirst}) and (\ref{acsecond}), we calculate the integral in eq. (\ref{finsol}) and find that in the rapid contraction regime, the low momentum modes behave as follows:
\be
\varphi^{(+)}=\frac{1}{(2\pi)^{3/2}\sqrt{2k}}\left(\frac{1}{a_{**}}+ika_{**}\left[\left(\frac{6c}{k^2}\right)^{1/4}\frac{1}{a_{**}^2}+\frac{t_M}{a_{**}^3}-\frac{2c}{a_{**}(t_M)}-\frac{2c}{a_{**}^2}\left(\frac{k^2}{6c}\right)^{1/4}-\frac{1}{a_*^2}\ln\left(\frac{\eta_*-\eta}{\eta_*}\right)\right]\right) \; .\nonumber
\ee
It follows from eqs. (\ref{sh1}), (\ref{sh5}) and the condition of smallness of momentum (\ref{midm}) that the main contribution comes from the first, constant term:
\be
\varphi^{(+)}=\frac{1}{(2\pi)^{3/2}\sqrt{2k}}\left(\frac{M_{pl}\sqrt{\lambda}}{a_{*}^2}\right)^{1/3} \; .\nonumber
\ee
With the standatd definition of the power spectrum,
\be
\langle\varphi(x,t)\varphi(x,t)\rangle=\int \frac{d^3k}{4\pi k^3}\mathcal{P_{\varphi}}(k) \; ,
\nonumber
\ee
we get
\be
\mathcal{P_{\varphi}}(k)=\frac{k^2}{4 \pi^2 a_{**}^2}=\frac{k^2}{4\pi^2}\left(\frac{M_{pl}\sqrt{\lambda}}{a_{*}^2}\right)^{2/3} \; .
\label{powspecf}
\ee
So, the power spectrum is blue, in agreement with Ref. \cite{hinterbicherkhouri}. According to eq. (\ref{svuaz}), the power spectrum of metric perturbations is
\be
\mathcal{P}_{h}=2\frac{32\pi}{M_{pl}^2}\mathcal{P_{\varphi}}=\frac{16k^2}{\pi}\left(\frac{\sqrt{\lambda}}{M_{pl}^2a_{*}^2}\right)^{2/3} \; ,
\nonumber
\ee
where the factor $2$ accounts for two polarizations.

Let us now turn to high momentum modes, such that $k\gg \sqrt{\lambda}M_{pl}a_{**}$, cf. eq. (\ref{midm}). These modes exit the horizon at the rapid contraction stage, so the field equation is
\be
\chi''+\frac{1}{4}\frac{\chi}{(\eta_*-\eta)^2}+k^2\chi=0 \; .
\nonumber
\ee
Its solution that tends to $e^{ik\eta}$ at large negative $\eta$ reads
\be
\chi^{(+)}=\sqrt{\frac{\pi k}{2}\left(\eta_*-\eta\right)}H_0^{(1)}\left[(\eta_*-\eta)k\right] \; .
\nonumber
\ee
After horizon crossing, this mode has constant part and logarithmically growing one. The dependence on $k$ is only logarithmic:
\be
\varphi^{(+)}=\frac{1}{(2\pi)^{3/2}\sqrt{2k}}\frac{\chi^{(+)}}{a}=\frac{\pi+2i\ln\left(\frac{k(\eta_*-\eta)}{2}\right)+2i\gamma}{4\sqrt{2}a_*\pi^2} \; .
\label{phisecond}
\ee
Hence, the power spectrum is again blue,
\be
\mathcal{P_{\varphi}}(k)=\frac{k^3}{8\pi a_{*}^2}\left\lbrace 1+\frac{4}{\pi^2}\left[\gamma+\ln\left(\frac{k(\eta_*-\eta)}{2}\right)\right]^2\right\rbrace \; .
\label{powspecs}
\ee

This power spectrum refers to time before the bounce. To obtain the power spectrum after the bounce, we match the solution (\ref{phisecond}), valid  before the bounce, to the general superhorizon solution after the bounce, i.e., at radiation domination, when $a(\eta)=\tilde{a}(\eta-\tilde{\eta})$. Using the instantaneous bounce approximation, cf. eqs. (\ref{Hbreh}) -- (\ref{razet}), we get
\be
\eta_b-\tilde{\eta}=\frac{a(\eta_b)}{a'(\eta_b)}=2(\eta_{*}-\eta_b)
\nonumber
\ee
The superhorizon solution to eq. (\ref{oburvo}) at radiation domination is
\be
\varphi^{(+)}=A+\frac{B}{\eta-\tilde{\eta}}
\nonumber
\ee
By matching this expression and its first derivative to eq. (\ref{phisecond}) at $\eta=\eta_b$, we get at late times
\be
\varphi^{(+)}=A=\frac{\pi+2i(\gamma-2)+2i\ln\frac{k(\eta_*-\eta_b)}{2}}{4\sqrt{2}a_{*}\pi^2}
\nonumber
\ee
So, the primordial power spectrum at radiation domination is
\be
\mathcal{P_{\varphi}}(k)=\frac{k^3}{8\pi a_{*}^2}\left\lbrace 1+\frac{4}{\pi^2}\left[\gamma-2+\ln\left(\frac{k(\eta_*-\eta_b)}{2}\right)\right]^2\right\rbrace \; ,
\label{powspecs2}
\ee
i.e.,
\be
\mathcal{P}_{h}=\frac{8k^3}{{M_{pl}^2 a_{*}^2}}\left\lbrace1+\frac{4}{\pi^2}\left[\gamma-2+\ln\left(\frac{k(\eta_*-\eta_b)}{2}\right)\right]^2\right\rbrace \; .
\nonumber
\ee

Let us now discuss possible effects of the tensor modes in the present and recent Universe. To this end, we again recall our assumption of the instantaneous bounce and defrosting. We make use of eq.(\ref{a**phys}) and write the present value of the critical momentum, eq.(\ref{midm}), as follows
\be
k_c=
\sqrt{\lambda}M_{pl}a_{**}=
T_0\left(\frac{M_{pl}\lambda}{T_{d}}\right)^{1/3}\left(\frac{2g_{*0}\cdot 1.66}{\sqrt{g_{*d}}}\right)^{1/3} \; ,
\nonumber
\ee
where the subscript $0$ refers to the present value, and we set $a_0=1$. We see that this momentum is fairly high for not too small $\lambda$.

We first consider direct search for gravity waves. Barring the very small value of $\lambda$, the relevant wavenumbers are below the critical value, $k<k_c$. Hence, these modes exit the horizon at the slow contraction stage, and we use eq. (\ref{powspecf}) for their power spectrum. We again recall eq. (\ref{a*phys}) and find for the energy density
\be
\frac{d\rho_{GW}}{d\ln k}=2\cdot(2\pi)^3k^2\mathcal{P}_{\varphi}(k)a_k^2=k^2\frac{T_{0}^2T_{d}^{2/3}\lambda^{1/3}}{M_{pl}^{2/3}}\frac{g_{*0}^{2/3}g_{*d}^{1/3}}{g_{*k}^{1/3}}\frac{(1.66)^{\frac{4}{3}}}{2^{\frac{2}{3}}} \; ,

\label{ogv}
\ee
where $a_{k}$ is the scale factor at horizon re-entry at radiation domination, $a_k=\frac{a_T^2H_T}{k}\left(\frac{g_{*T}}{g_{*T_k}}\right)^{1/6}$. The result (\ref{ogv}) is hopelessly small: as an example, for $k\sim 100~\mbox{Hz}$ we find that even with $T_d\sim M_{pl}$ and $\lambda\sim 1$
\be
\frac{d\Omega_{GW}}{d\ln k}\simeq  5\times 10^{-24} \; , 
\nonumber
\ee
which is far below the projected sensitivity of gravity wave detectors.

We now turn to the BBN constraint
\be
\rho_{GW}(t_{NS})\lesssim\frac{\pi^2}{30}T_{NS}^4  \; .
\label{BBNn1}
\ee
We consider the case in which the rapid contraction regime does occur at the contracting stage (in the opposite case one obtains even weaker constraint on the parameters of the model). Since the power spectrum is blue, we make use of the result (\ref{powspecs2}) and obtain
\be
\rho_{GW}(t_{NS})=\int\limits^{k_{max}}d^3k~2\cdot\mathcal{P}_{\phi}(k)\left(\frac{a_k}{a_{NS}}\right)^2\frac{2\pi^2}{a_{NS}^2k}\simeq2\pi^2(1.66)^4
\frac{T_{d}^4T_{NS}^4}{M_{pl}^4}g_{*NS}^{4/3}g_{*d}^{2/3}
\nonumber
\ee
where $k_{max}$ is the highest momentum of modes that ever exit the horizon, $k_{max}/a_b\sim H_b$, or $k_{max}(\eta_*-\eta_b)\sim 1$. Thus, the constraint (\ref{BBNn1}) translates into a weak bound on the defrosting temperature
\be
T_{d}\lesssim 0.22\cdot\frac{M_{pl}}{g_{*NS}^{1/3}g_{*d}^{1/6}} \; .
\nonumber
\ee
Thus, the pseudo-conformal Universe model is viable in a wide range of its parameters.

\section{Conclusion}

To summarize, we have solved the classical equations of motion and found the evolution of the Universe in the pseudo-conformal model. There are two regimes, one after another. The first is ekpyrotic (slow contraction) and the second is fairly rapid collapse. We then calculated the tensor spectra generated in the two regimes. We have seen that in both regimes there is no considerable production of tensor modes, hence this model is acceptable from the viewpoint of generating the cosmological perturbations.
\section{Acknowledgements}

The author is grateful to Valery Rubakov for valuable discussions and interest in this work. This work has been supported in part by RFBR grant 12-02-00653 and by RFBR grant for young scientists 12-02-31778, the Federal Agency for Science and Innovations under state contract 16.740.11.0583, the grant of the President of the Russian Federation NS-5590.2012.2 and the Ministry of Education and Science contract 8412. The author acknowledges the support by the Dynasty Foundation.

\section*{Appendix}

To cross check the calculation leading to the result (\ref{finsol0}), we explicitly solve eq. (\ref{eqlt}) in the limit $\frac{c}{\eta^2}\ll 1$. We write
\be
\chi=e^{ik\eta}(1+f(\eta,k)) \; .
\nonumber
\ee
Then eq. (\ref{eqlt}) takes the following form:
\be
f''+2ikf'+\frac{6c}{\eta^4}(1+f)=0 \; .
\label{crchurv}
\ee
For $c/\eta^2\ll 1$, the correction is small, $f\ll 1$, so that eq. (\ref{crchurv}) is
\be
f''+2ikf'+\frac{6c}{\eta^4}=0 \; ,
\nonumber
\ee
The solution to this equation, which asymptotes $f=0$ as $\eta\rightarrow -\infty$, is
\be
f=-\int\limits_{-\infty}^{\eta}e^{-2ik\eta'} \int\limits_{-\infty}^{\eta'}\frac{6c}{\eta''^4}e^{2ik\eta''}d\eta'' d\eta' \; .
\nonumber
\ee
Upon changing the order of integration we get
\be
f=-
\int\limits_{-\infty}^{\eta}\frac{6c}{\eta'^4}\left(e^{-2ik(\eta-\eta')}-1\right)d\eta' \; ,
\nonumber
\ee
and after three integrations by parts we come to
\be
f=-\frac{c}{\eta^2}-\frac{2ikc}{\eta}+\int\limits_{-\infty}^{\eta}\frac{c}{\eta'}e^{-2ik(\eta-\eta')}(2ik)^2d\eta' \; .
\nonumber
\ee
This means that in the limit $k\eta\ll 1$ the solution is
\be
\chi=e^{ik\eta}\left(1-\frac{c}{\eta^2}-\frac{2ikc}{\eta}+...\right)=1-\frac{c}{\eta^2}-\frac{3ikc}{\eta}+ik\eta+...~~ \; .
\nonumber
\ee
This confirms our result (\ref{finsol0}).


\begin{thebibliography}{21}
\bibitem{ekpyrosis}
J. Khoury, B. Ovrut, P. Steinhardt, N. Turok, "The ekpyrotic Universe: colliding branes and the origin of the hot Big Bang", Phys.\ Rev.\ D {\bf 64} 123522 (2001) arXiv:hep-th/0103239\\
\\
J. Khoury, B. Ovrut, N. Seiberg, P. Steinhardt, N. Turok, "From Big Crunch to Big Bang", Phys.\ Rev.\ D {\bf 65} 086007 (2002) arXiv:hep-th/0108187

\bibitem{ekpnew}
J. Khoury, P. Steinhardt, "Adiabatic Ekpyrosis: Scale-Invariant Curvature Perturbations from a Single Scalar Field in a Contracting Universe", Phys.\ Rev.\ Lett.\  {\bf 104}, 091301 (2010), arXiv:0910.2230\\
\\
D. Lyth, "The primordial curvature perturbation in the ekpyrotic Universe", Phys.\ Lett.\ B {\bf 524}, 1 (2002), arXiv:hep-ph/0106153\\
\\
R. Brandenberger, F. Finelli, "On the Spectrum of Fluctuations in an Effective Field Theory of the Ekpyrotic Universe", JHEP {\bf 0111}, 056 (2001), arXiv:hep-th/0109004\\
\\
J. Khoury, B. Ovrut, P. Steinhardt, N. Turok, "Density Perturbations in the Ekpyrotic Scenario", Phys.\ Rev.\ D {\bf 66} 046005 (2002), arXiv:hep-th/0109050\\
\\
P. Creminelli, A. Nicolis, M. Zaldarriaga, "Perturbations in bouncing cosmologies: dynamical attractor vs scale invariance", Phys.\ Rev.\ D {\bf 71} 063505 (2005), arXiv:hep-th/0411270\\
\\
Y. Cai, D. Easson, R. Brandenberger, "Towards a nonsingular bouncing cosmology", JCAP {\bf 1208}, 020 (2012), arXiv:1206.2382

\bibitem{hinterbicherkhouri}
K. Hinterbichler, J. Khoury, "The pseudo-conformal Universe: scale invariance from spontaneous breaking of conformal symmetry", JCAP {\bf 1204}, 023 (2012),  arXiv:1106.1428

\bibitem{hinterbicherkhourilr}
M. Libanov, V. Rubakov, "Dynamical vs spectator models of (pseudo-)conformal Universe", Teor. Mat. Fiz. {\bf 173}, 149 (2012), Theor.\ Math.\ Phys.\ {\bf 173}, 1457 (2012), arXiv:1107.1036 \\


\bibitem{confinfl1}
I. Antoniadis, P. Mazur and E. Mottola, "Conformal invariance and cosmic background radiation", Phys.\ Rev.\ Lett.\  {\bf 79}, 14 (1997), arXiv:astro-ph/9611208.

\bibitem{confinfl2}
V. Rubakov, "Harrison-Zeldovich spectrum from conformal invariance", JCAP {\bf 0909}, 030 (2009), arXiv:0906.3693

\bibitem{confinfl3}
P. Creminelli, A. Nicolis and E. Trincherini, "Galilean Genesis: an alternative to inflation", JCAP {\bf 1011}, 021 (2010); arXiv:1007.0027.

\bibitem{confnew}
M. Libanov, V. Rubakov, "Cosmological density perturbations from conformal scalar field: infrared properties and statistical anisotropy", JCAP {\bf 1011}, 045 (2010),2010, arXiv:1007.4949 \\
\\
M. Libanov, S. Mironov, V. Rubakov, "Properties of scalar perturbations generated by conformal scalar field", Prog.\ Theor.\ Phys.\ Suppl.\  {\bf 190}, 120 (2011), arXiv:1012.5737 \\
\\
M. Libanov, S. Ramazanov, V. Rubakov, "Scalar perturbations in conformal rolling scenario with intermediate stage", JCAP {\bf 1106}, 010 (2011), arXiv:1102.1390 \\
\\
M. Libanov, S. Mironov, V. Rubakov, "Non-Gaussianity of scalar perturbations generated by conformal mechanisms", Phys.\ Rev.\ D {\bf 84}, 083502 (2011), arXiv:1105.6230 \\

\bibitem{curv}
 A. Linde and V. Mukhanov,
  "Nongaussian isocurvature perturbations from inflation",
Phys.\ Rev.\ D {\bf 56}, 535 (1997), astro-ph/9610219.\\
\\
  K. Enqvist and M. Sloth,
  "Adiabatic CMB perturbations in pre - big bang string cosmology",
Nucl.\ Phys.\ B {\bf 626}, 395 (2002), hep-ph/0109214.\\
\\
D. Lyth and D. Wands,
  "Generating the curvature perturbation without an inflaton",
Phys.\ Lett.\ B {\bf 524}, 5 (2002), hep-ph/0110002.\\
\\
T. Moroi and T. Takahashi,
  "Effects of cosmological moduli fields on cosmic microwave background",
Phys.\ Lett.\ B {\bf 522}, 215 (2001), Erratum-ibid.\ B {\bf 539}, 303 (2002), hep-ph/0110096.\\
\\
K. Dimopoulos, D. Lyth, A. Notari and A. Riotto,
  "The Curvaton as a pseudoNambu-Goldstone boson",
JHEP {\bf 0307}, 053 (2003), hep-ph/0304050.


\bibitem{moddec}
  G. Dvali, A. Gruzinov and M. Zaldarriaga,
  "A new mechanism for generating density perturbations from inflation",
Phys.\ Rev.\ D {\bf 69}, 023505 (2004), astro-ph/0303591.\\
\\
 L. Kofman,
  "Probing string theory with modulated cosmological fluctuations", astro-ph/0303614.\\
\\
 G. Dvali, A. Gruzinov and M. Zaldarriaga,
  "Cosmological perturbations from inhomogeneous reheating, freezeout, and mass domination",
Phys.\ Rev.\ D {\bf 69}, 083505 (2004), astro-ph/0305548.


\bibitem{wandbrand}
Y. Wang and R. Brandenberger,
  "Scale-Invariant Fluctuations from Galilean Genesis",
arXiv:1206.4309.

\bibitem{gravwave}
L. Boyle, P. Steinhardt, N. Turok, "The Cosmic Gravitational-Wave Background in a Cyclic Universe", Phys.\ Rev.\ D {\bf 69}, 127302 (2004) arXiv:hep-th/0307170

\bibitem{def}
  L. Levasseur Perreault, R. Brandenberger and A. Davis,
  "Defrosting in an Emergent Galileon Cosmology",
Phys.\ Rev.\ D {\bf 84}, 103512 (2011),
arXiv:1105.5649.

\end{thebibliography}
\end{document}